\documentclass[11pt,twoside]{article}

\usepackage{asp2006}
\usepackage{epsf}
\usepackage{psfig}
\usepackage{lscape}
\usepackage{graphicx}
\usepackage[breaklinks=true]{hyperref}

\usepackage{ulem}

\usepackage{amsmath}

\usepackage{amssymb}

\def\thebibliography#1{\small\section*{\small References}
\list{\null}{\leftmargin 3em\labelwidth 0pt\labelsep 0pt\itemindent -3em
\itemsep 0pt \parsep 0pt\usecounter{enumi}}
\def\refpar{\relax}
\def\newblock{\hskip .11em plus .33em minus .07em}
\sloppy\clubpenalty4000\widowpenalty4000
\sfcode`\.=1000\relax}

\begin{document}
\title{Numerical Studies of Relativistic Jets}   
\author{Ramesh Narayan, Alexander Tchekhovskoy}   
\affil{Harvard-Smithsonian Center for Astrophysics, 60 Garden Street, Cambridge, MA 02138, USA}    
\author{Jonathan McKinney}   
\affil{Kavli Institute for Particle Astrophysics and Cosmology, Stanford
University, Stanford, CA 94305, USA}    

\begin{abstract} 

The connection between collimation and acceleration of magnetized
relativistic jets is discussed.  The focus is on recent numerical
simulations which shed light on some longstanding problems.

\end{abstract}


\section{Introduction}   

Understanding the physics of relativistic jets, notably their
acceleration and collimation, has been a major area of study for
several decades.  It is widely agreed that jets are a magnetic
phenomenon.  The basic paradigm involves a bundle of field lines that
is attached at its base to a central compact star -- black hole (BH)
or neutron star (NS) -- or to an accretion disk.
Rotation of the star and/or disk causes a helical outgoing
magnetohydrodynamic (MHD) wave which accelerates any gas that is
frozen into the field lines.

A great deal of analytical work has been done on jets, and the reader
is referred to \citet{begelman_asymptotic_1994},
\citet{vla03a,vla03b}, \citet{bes06}, \citet{nar07},
\citet{lyub09,lyub09b}, and references therein, for an introduction to
the vast literature.  However, the equations describing MHD jets are
nonlinear and the problem is complicated, so there is a limit to what
can be accomplished purely analytically.  Fortunately, in recent
years, numerical investigations have begun to contribute to the field,
facilitated by the development of robust relativistic MHD codes
\citep{koi00,kom01,mck04,dhkh05,mck07a,mck07b,kom07,kom09,tch08,tch09,tch09c,tch09b}.

\section{Force-Free Jets}

Before considering the full MHD problem, it is useful to focus first
on force-free jets.  The force-free approximation is a simplification
of ideal MHD in which we include electric and magnetic fields, and the
corresponding charges and currents, but we ignore the inertia of the
gas.  The problem thus reduces to pure electrodynamics in a perfectly
conducting medium.  This very simple approximation provides
surprisingly useful insight into the general problem.

\subsection{Collimation}
\label{sec_collimation}

\citet{mic73} derived an analytical solution for a force-free wind
from a rotating star.  He assumed a split monopole configuration in
which the magnetic field is radial and has a constant strength at the
surface of the star, with lines pointed into the star over one
hemisphere and out over the other hemisphere.

The rotation of the star causes the magnetic field to develop a
toroidal component $B_\phi$ which dominates with increasing distance:
$B_\phi = -(\Omega R/c)B_p$, where $\Omega$ is the angular velocity of
the star, $R$ is cylindrical radius with respect to the rotation axis,
and $B_p$ is the field strength in the poloidal ($r\theta$) plane.
However, even though $B_\phi$ dominates over $B_p$ at large radii
($-B_\phi/B_p = \Omega R/c \gg 1$), nevertheless there is no change in
the poloidal geometry -- the poloidal field continues to remain
radial.  This is surprising since the tension associated with toroidal
field curvature, the so-called hoop stress, is very large.  Why does
this not collimate the jet?  It turns out that hoop stress is pefectly
canceled by an electric force in the opposite direction, and hence
there is no net tendency to collimate the flow.  (This is a purely
relativistic phenomenon -- the electric field is negligible in
nonrelativistic MHD.)

A similar result is seen in the paraboloidal force-free solution
derived by \citet{blandford_accretion_disk_electrodynamics_1976} and
\citet{bz77}, and a more general class of self-similar force-free
solutions studied by \citet{nar07}.  Although hoop stress is very
large at large distances from the base of the jet, once again it has
small or no effect on the collimation of the jet.

All this suggests that a relativistic jet is unlikely to
self-collimate.  It needs to be collimated by an external agency.
Assuming that the jets in active galactic nuclei (AGN) and X-ray
binaries (XRBs) are associated with field lines from a spinning
accreting BH, these jets must be confined by a wind from the
surrounding accretion disk.  In the collapsar model of gamma-ray
bursts (GRBs), the confinement is presumably due to the stellar
envelope of the collapsing star.  If no confining medium is present,
as in the case of relativistic outflows from radio pulsars, the flow
is nearly radial.

\subsection{Acceleration}

Force-free jets accelerate very efficiently.  In the simplest
geometries, viz., split monopole and paraboloidal, the Lorentz factor
of the outflow varies asymptotically as $\gamma \approx \Omega R/c$,
increasing linearly with cylindrical radius.  In the case of a
paraboloidal jet, $\gamma\propto z^{1/2}$ (since $R\propto z^{1/2}$),
where $z$ is distance along the jet.

For more general geometries such as $R \propto z^{(2-\nu)/2}$
\citep{nar07}, $\gamma$ depends both on the value of $\Omega R/c$ and
on the poloidal curvature of field lines.  As a result, the run of
$\gamma$ with $z$ exhibits two distinct asymptotic regimes (see
\citealt{tch08} for details).  Nevertheless, at any given $z$, the
field line with the largest Lorentz factor still satisfies
$\gamma\propto z^{1/2}$.  Thus, this simple and convenient scaling
appears to be fairly general.

\subsection{Jet Power}

Since a force-free jet is purely electromagnetic, the energy flux is
given by the Poynting vector $\vec{E}\times\vec{B}/4\pi$ (we set
$c=1$).  For a rotating axisymmetric jet, $E = (\Omega R/c)B_p$ and is
pointed perpendicular to the poloidal field line in the $(r\theta)$
plane.  In a relativistic jet electric and magnetic fields are
nearly equal in magnitude (\S\ref{sec_collimation}), 
$E\approx\lvert B_\phi\rvert$,  therefore
the poloidal component of the energy flux is given by
$E\lvert B_\phi\rvert/4\pi = (\Omega R/c)^2 B_p^2/4\pi$.  Applying this result at
the BH horizon, the total electromagnetic power flowing out of a
spinning BH is given by $P_{\rm jet} = k \Phi_{\rm tot}^2\Omega_H^2$,
where $\Phi_{\rm tot}$ is the total magnetic flux threading the
horizon, $\Omega_H$ is the angular frequency of the spacetime at the
horizon, and $k$ is a constant which depends on the field geometry
(see \citealt{tch09c} for details).  Although this result for $P_{\rm
jet}$ is derived for a force-free jet in the limit of a slowly
spinning BH \citep{bz77}, it turns out to be surprisingly accurate
even for rapidly spinning BHs, and also for MHD jets \citep{tch09c}.

One interesting question is whether the $\Omega_H^2$ scaling of power
is sufficient to explain the radio loud/quiet dichotomy of AGN.
Observations indicate a factor of 1000 difference in radio power
between radio loud and quiet AGN \citep{ssl07}.  Can this be entirely
due to differences in the BH spins of the underlying populations?
\citet{tch09c} show that a model in which the BH is surrounded by a
thick accretion disk, such that the jet subtends only a narrow solid
angle around the axis, has a steep variation of jet power with spin --
$P_{\rm jet}\propto \Omega_H^4$ or even $\Omega_H^6$ (a similar
steep dependence was observed by \citealt{mck05} in the simulations of BHs
with thick turbulent tori).  Such a model is
perfectly compatible with the observed radio loud/quiet dichotomy,
whereas a BH with a thin disk around it would have $P_{\rm
jet}\propto\Omega_H^2$ and is not likely to have a large range of jet
power.  This explanation of the radio loud/quiet dichotomy requires
AGN jets, especially in low-luminosity systems, to be associated with
thick advection-dominated accretion flows (ADAFs).  There is, in fact,
independent evidence for such a connection \citep{nm08}.

\subsection{Stability}

According to the Kruskal-Shafronov criterion (e.g., \citealt{bat78}),
cylindrical MHD configurations in which the toroidal field dominates
are violently unstable to the $m=1$ kink (or screw) instability.
Since all models of relativistic magnetized jets have $B_\phi \gg
B_p$, jets ought to be highly unstable.  This has been recently
confirmed by numerical simulations~\citep{mizuno2009}.  
Why are the jets observed in
Nature coherent over enormous length scales?

\citet*{nlt09} studied the stability of force-free jets and found that
the growth rate of the kink mode is quite low.  There are two reasons
for this.  First, although jets may have $B_\phi \gg B_p$ in the ``lab
frame,'' they typically have $B_\phi \sim B_p$ in the comoving frame
of the fluid.  This makes the instability less severe.  Second, as a
result of time dilation, the growth rate as measured in the lab frame
is further suppressed by a factor of $1/\gamma$.  Note that the above
study was limited to a very simple force-free model.  The stability of
more realistic relativistic MHD jets is an open question.

\begin{figure}
\plottwo{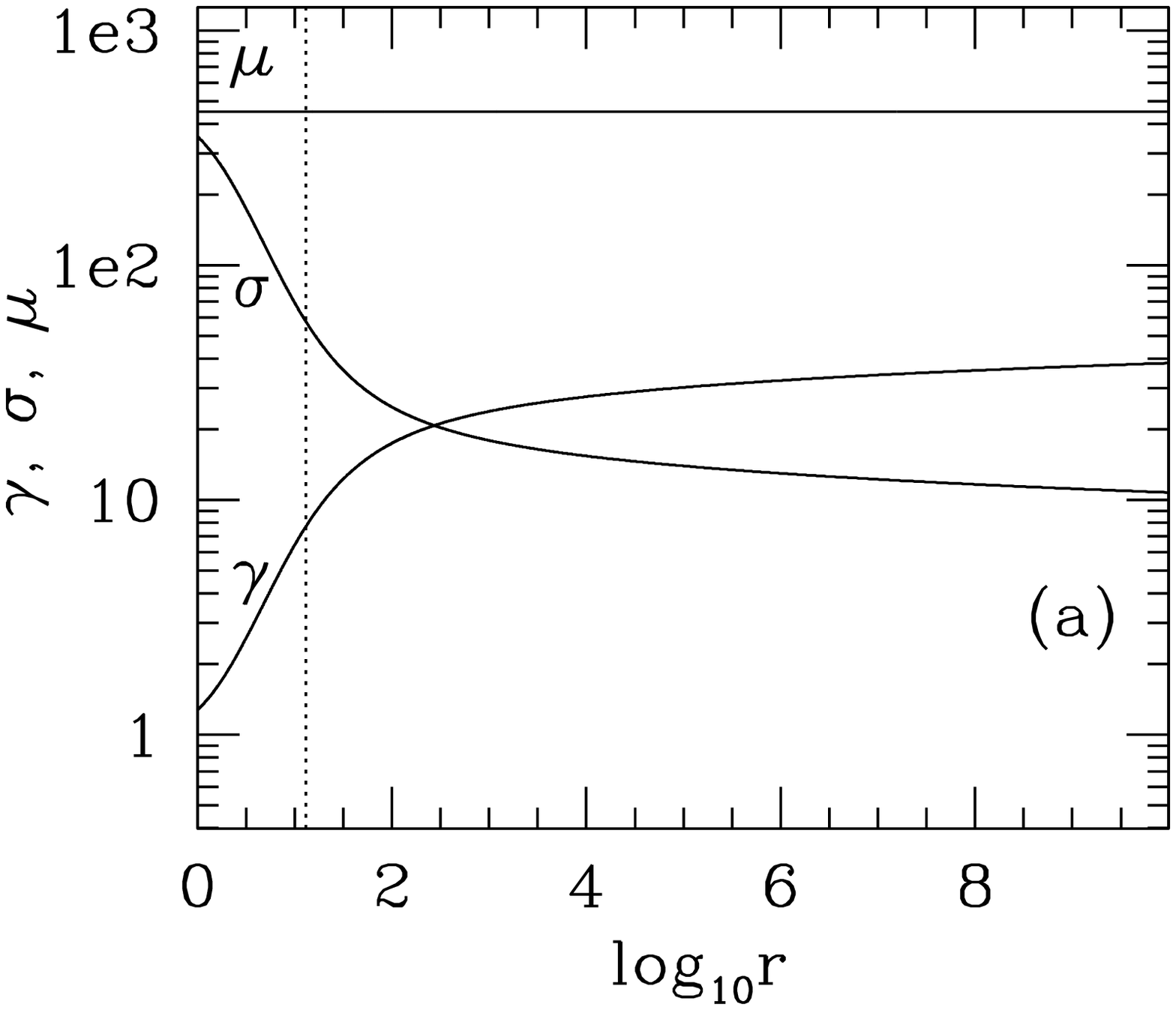}{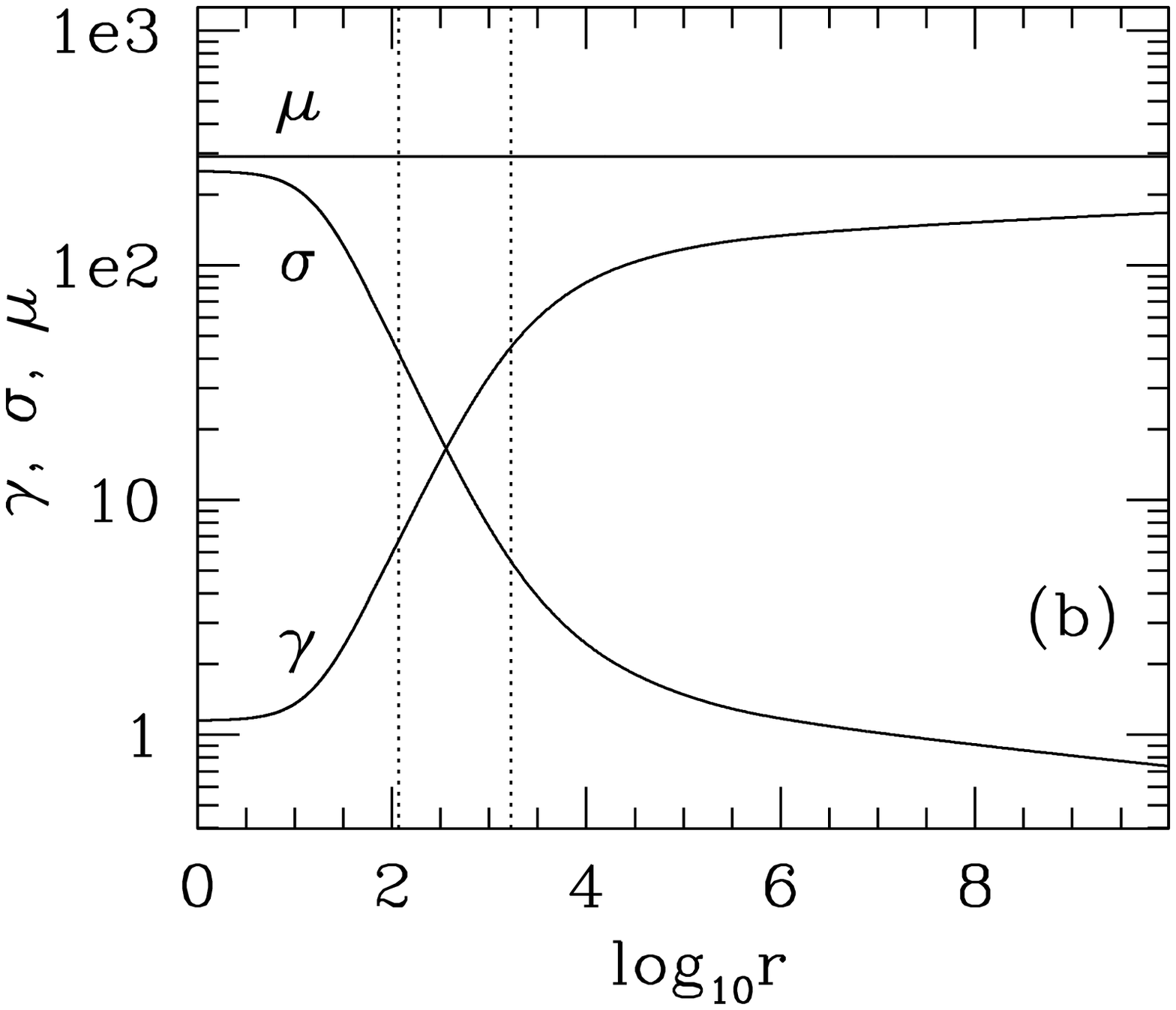}
\caption{[Panel (a)] Solid lines show Lorentz factor $\gamma$,
  magnetization parameter $\sigma$ and conserved energy flux $\mu\
  (=450$) as a function of distance $r$ (roughly in units of the BH
  horizon radius) along an equatorial field line ($\theta=90^\circ$)
  for a mass-loaded split-monopole outflow.  Initially, $\gamma$
  increases rapidly with $r$.  However, beyond the fast point
  (indicated by the vertical dotted line), it grows only
  logarithmically.  As a result, even at astrophysically relevant
  distances, $\gamma$ is only $\sim40$ and most of the energy remains
  locked up in the magnetic field: $\sigma\sim10$.  This is the
  $\sigma$ problem.  [Panel (b)] Similar to panel (a), but for a polar
  field line at $\theta=10^\circ$.  The two vertical dotted lines show
  the positions of the fast point (left) and the causality point
  (right).  Here the flow continues to accelerate beyond the fast
  point and up to the causality point.  As a result, it is less
  affected by the $\sigma$ problem and achieves a larger final Lorentz
  factor $\gamma_{\rm final}\sim\mu$.  (Based on \citealt{tch09})}
\label{fig1}
\end{figure}

\section{Relativistic MHD Jets}

We now consider MHD jets with gas inertia, though, for simplicity, we
ignore gas pressure.  A steady axisymmetric MHD flow has a number of
conserved quantities: (i) The angular velocity $\Omega$ is constant
along any field line.  (ii) The net mass flowing along a bundle of
field lines is conserved, which means that the quantity $\gamma\rho
v_p/B_p$ is constant along each field line, where $\rho$ is the mass
density in the rest frame of the gas and $v_p$ is the poloidal
$3$-velocity of the gas in the lab frame.  (iii) The angular momentum
flowing along the bundle of lines is conserved.  (iv) Finally, the
energy flowing along the bundle is also conserved, which implies that
\begin{equation}
\mu=\left\{\lvert\vec{E}\times\vec{B}_\phi/4\pi\rvert+(\gamma^2\rho v_p)\right\}
/\gamma\rho v_p
=\,{\rm constant},
\end{equation}
where the first term in the numerator is the Poynting energy flux and
the second is the gas kinetic energy flux.  This equation may be
rewritten as
\begin{equation}
\mu=\gamma(\sigma+1), \qquad \sigma = \lvert\vec{E}\times\vec{B}_\phi/4\pi\rvert/
(\gamma^2\rho v_p),
\label{musigma}
\end{equation}
where the magnetization parameter $\sigma$ represents the ratio of
electromagnetic to kinetic energy flux at any point in the flow.

\subsection{Acceleration and the $\sigma$ Problem}

Equation (\ref{musigma}) has a simple interpretation.  At the base of
the jet, before the gas has accelerated, we have $\gamma\approx1$, and
so the energy flow is almost entirely in the form of Poynting flux:
$\sigma\gg1$, $\mu\gg 1$.  Here the jet is magnetically very dominated
and it behaves for all intents like a force-free jet.  As the jet
moves out, $\gamma$ increases and the kinetic energy flux grows at the
expense of the Poynting flux.  Thus $\sigma$ decreases as $\gamma$
increases, but in such a way as to keep $\mu$ constant.  If the
acceleration is efficient, then we expect $\sigma$ to become
vanishingly small at large distance and all the initial energy to end
up in the gas; we would then have $\gamma = \gamma_{\rm max}=\mu$.  In
fact, this ideal is rarely achieved.

The problem is most clearly seen in the split monopole geometry,
where analytical results are available for equatorial field lines
\citep{mic69,bes98,tch09}.  The flow accelerates initially just as in
the force-free case, with $\gamma \sim \Omega R/c$.  However, once the
flow speed exceeds the fast magnetosonic wave speed in the medium,
which happens when $\gamma \sim \mu^{1/3}$, or equivalently when
$\gamma\sim\sigma^{1/2}$ (from eq. \ref{musigma} with $\mu\gg1$), the
gas loses contact with gas behind it and acceleration slows down
drastically.  Beyond this point, there is only a growth in $\gamma$ of
order a logarithmic factor, so the final $\gamma$ is given by
$\gamma_{\rm final}\sim C\sigma^{1/2}$, where $C$ is a logarithm.  If
$\mu$ is large, as it must be for a highly relativistic flow, the
asymptotic value of $\sigma$ remains large and most of the
energy is still carried in the form of Poynting flux rather than as
gas kinetic energy.  This is the $\sigma$ problem.

The $\sigma$ problem is illustrated by a numerical example in
Fig. \ref{fig1}a.  As we see, the problem is not a lack of energy.  In
this example, $\mu=450$ which means there is plenty of energy
available and the gas could in principle accelerate up to $\gamma_{\rm
max}\sim450$.  However, it only accelerates up to $\gamma_{\rm
final}\simeq 40$ because the energy remains stuck in the magnetic
field.

Simulations such as those shown in Figs. \ref{fig1}--\ref{fig2} of this paper are
challenging and have only recently become possible with the use 
of the general relativistic MHD code HARM \citep{gam03}, including recent improvements
(\citealt{tch07,tch08}). There are two
difficulties with these calculations.  First, the large value of
$\sigma$ means that the electromagnetic and kinetic terms in the
energy equation are very different in magnitude.  The equations are
thus stiff, making it very difficult to maintain accuracy.  Second,
magnetic acceleration is relatively slow: $\gamma$ varies only as
$z^{1/2}$ even when acceleration is efficient, and it grows
logarithmically when it is inefficient.  Therefore, in order to obtain
useful results, it is necessary to simulate relativistic jets over
very large length scales, which is obviously challenging.  As of this
writing, only one other group (Komissarov and collaborators) is able
to carry out such calculations.

\subsection{No $\sigma$ Problem for Collimated Jets!}

Observations suggest that relativistic jets in Nature do not suffer
from the $\sigma$ problem.  \citet{lind1989} showed that the
interaction of a jet with an external medium is very different
depending on whether $\sigma\ll1$ or $\sigma\gg1$.  In the former case
there is a clear hot spot (or working surface) where the jet shocks
with the medium, and there is a well-defined backflow in a cocoon.
This is similar to what is observed in FR II jets.  In contrast, a
magnetically dominated jet drills through the medium and has hardly
any cocoon.  It appears that AGN jets manage to achieve small values
of $\sigma$ before they shock on the external medium.  Why are they
not limited by the $\sigma$ problem?

A key clue was provided by \citet{tch09} who showed that equatorial
and polar field lines in the split monopole geometry behave very
differently.  This is illustrated in Fig. \ref{fig1}b which shows that
acceleration along a polar field line continues well past the fast
magnetosonic point.  In fact, acceleration slows down only after the
gas crosses another critical point, the ``causality point,'' beyond
which it can no longer communicate with the axis.  The introduction of
this new critical point means that the final Lorentz factor of the gas
is a function of the angle $\theta$ between the poloidal field and the
axis:
\begin{equation}
\gamma_{\rm final}\sin\theta \sim C \sigma^{1/2},
\label{gammafinal}
\end{equation}
where $C$ is the same logarithmic factor as before.  The smaller the
value of $\theta$, the larger the $\gamma_{\rm final}$ that the gas
can achieve for a given value of $\mu$.  Thus, flows that remain close
to the axis accelerate more efficiently.

\begin{figure}[t]
\plotone{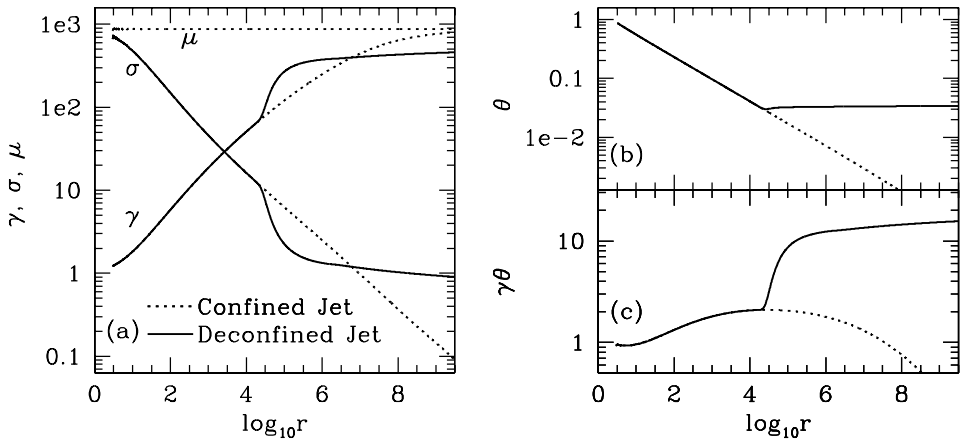}
\caption{The dotted lines in panel (a) show the variation of $\gamma$,
  $\sigma$ and $\mu\ (\approx900)$ as a function of distance $r$ (BH
  units) for an MHD jet that is confined by a rigid wall with a shape
  given by $R \propto z^{5/8}$.  As for all externally collimated
  jets, we see that this model does not suffer from the $\sigma$
  problem.  Asymptotically, the jet achieves a large Lorentz factor
  $\gamma\approx\mu$ and a low magnetization $\sigma\ll1$.  The dotted
  line in panel (b) shows the variation of the jet opening angle
  $\theta$, and in panel (c) the product $\gamma\theta$.  Note that
  $\gamma\theta\lesssim1$, which is a general feature of fully
  collimated jets. Solid lines in the three panels correspond to a jet
  that is confined out to a distance $r=10^{4.5}$ and is then allowed
  to move freely in vacuum.  The jet undergoes a period of rapid
  acceleration just as it exits into vacuum (see panel a), but its
  opening angle does not change (panel b).  As a result,
  asymptotically, the jet has nearly the same Lorentz factor as in the
  previous example, $\gamma \sim \mu/2$, $\sigma\sim1$, but now
  $\gamma\theta\sim15$ (panel c).  This model may be relevant for
  understanding GRB jets.  (Based on \citealt{tch09b})}
\label{fig2}
\end{figure}

This result is confirmed by numerical simulations of paraboloidal and
other types of confined jets.  The dotted lines in Fig. \ref{fig2}
show an example of a confined jet with $\mu\sim10^3$.  This jet
accelerates with no trouble up to $\gamma_{\rm final}$ of nearly
$10^3$, achieving an asymptotic $\sigma<0.1$.  The lesson from this
example is that, so long as a jet is provided adequate collimation, it
will accelerate smoothly without being limited by the $\sigma$
problem.

\subsection{How are $\gamma$ and $\theta$ related?}

One additional interesting result is seen in Fig. \ref{fig2}: while
$\gamma$ increases, the jet angle $\theta$ simultaneously decreases.
In fact, it appears that $\gamma\theta\sim1$, as noted by
\citet{kom09}.  Do relativistic jets always satisfy this constraint?
This can be tested with GRB jets, where observations provide strong
constraints on both $\gamma$ and $\theta$ (e.g., \citealt{pir05}).

Long duration GRBs typically accelerate to $\gamma \sim$ few hundred
before decelerating in their afterglow phase.  At the same time, jet
breaks observed in GRB afterglows indicate typical jet opening angles
$\theta\sim0.05$ radian.  Thus, GRB jets apparently have
$\gamma\theta\sim$ few tens.  This appears to violate the
\citet{kom09} relation mentioned above, which is also confirmed by
Fig. \ref{fig2}.

Fortunately, there is a simple solution.  The solid lines in
Fig. \ref{fig2} show a second simulation in which the jet is initially
confined over a few decades in distance and it is then allowed to move
freely in vacuum.  This is meant to mimic a collapsar in which the jet
is confined so long as it is inside the stellar envelope, but becomes
free once it escapes from the star.  In this numerical example, we see
that the gas experiences a burst of acceleration just as it is
deconfined, without changing its opening angle very much.
Asymptotically, the jet satisfies the scaling given in equation
(\ref{gammafinal}) with $C\sim 10$.  The particular simulation shown
here has $\gamma_{\rm final}\sim500$, $\theta_{\rm final}\sim0.04$,
$\sigma_{\rm final}\sim1$, which is close to typical values observed
in GRBs.  (Note that $\sigma_{\rm final}$ is not directly measured.
However, the fact that GRBs emit a substantial fraction of their
enerrgy in prompt emission implies that their jets cannot be
Poynting-dominated.)

There is room for further work in this area.  For instance, if
observations by the Fermi Observatory routinely find $\gamma_{\rm
final}>10^3$ \citep{kd09} and $\theta_{\rm final} > 0.1$
\citep{cenko_grb_breaks_2009} for many GRBs, we
would need to explain why these jets have $\gamma\theta>100$.  The
constant $C$ in equation (\ref{gammafinal}) is a logarithmic factor
and is unlikely to be much larger than about $10-20$.  Therefore,
either we must accept that $\sigma\gg1$, i.e., GRB jets are
Poynting-dominated and somehow manage to convert a large fraction of
their energy to promt gamma-rays, or that MHD is not the appropriate
framework for understanding GRB jets.

\section{Conclusion}

The main message of the work described here is that collimation is the
key to relativistic jets.  If a jet is suitably confined by an
external medium, it will accelerate without difficulty.  While GRB
jets have a natural collimating medium in the surrounding stellar
envelope, the situation is less obvious for jets in AGN and XRBs.  A
disk wind is the most likely collimator in these systems, which
suggests that many prominent jets are probably associated with
geometrically thick accretion disks with strong winds and outflows,
i.e., ADAFs of various kinds \citep{nm08}.  This connection is worth
exploring further.

Equation (\ref{musigma}) shows that the maximum Lorentz factor that a
jet can achieve is equal to $\mu$.  A highly relativistic jet requires
a large value of $\mu$, which means that the jet must start out very
magnetically dominated at its base with negligible mass-loading.
There is very little understanding of how mass-loading works.
Qualitatively, one imagines that field lines that are connected to the
disk have a ready supply of disk gas for mass-loading, whereas lines
that penetrate the BH horizon are more likely to be mass-free.  One
thus suspects that the most extreme jets probably emerge from spinning
BHs rather than from disks.

Finally, in all of our discussion we assumed that a coherent magnetic
field is already present in the system.  The origin of this field is
not understood.  It could be advected in by the accretion disk from
outside.  However, the efficiency of such advection is poorly
understood and the topic is controversial.

\acknowledgements 

This work was supported in part by NASA grant NNX08AH32G and NSF grant
NSF AST-0805832.  The simulations were 
run on the Odyssey cluster supported by the Harvard FAS Research
Computing Group and on the TeraGrid resources~\citep{catlett2007tao}
provided by the Louisiana Optical Network Initiative
(\href{http://www.loni.org}{http://www.loni.org}) with the support of
NSF.




\begin{thebibliography}{37}
\expandafter\ifx\csname natexlab\endcsname\relax\def\natexlab#1{#1}\fi

\bibitem[{{Bateman}(1978)}]{bat78}
{Bateman}, G. 1978, {MHD instabilities} (Cambridge, Mass., MIT Press, 1978.~270
  p.)

\bibitem[{Begelman \& Li(1994)}]{begelman_asymptotic_1994}
Begelman, M.~C. \& Li, Z.-Y. 1994, \apj, 426, 269

\bibitem[{{Beskin} {et~al.}(1998){Beskin}, {Kuznetsova}, \& {Rafikov}}]{bes98}
{Beskin}, V.~S., {Kuznetsova}, I.~V., \& {Rafikov}, R.~R. 1998, \mnras, 299,
  341

\bibitem[{{Beskin} \& {Nokhrina}(2006)}]{bes06}
{Beskin}, V.~S. \& {Nokhrina}, E.~E. 2006, \mnras, 367, 375

\bibitem[{{Blandford}(1976)}]{blandford_accretion_disk_electrodynamics_1976}
{Blandford}, R.~D. 1976, \mnras, 176, 465

\bibitem[{{Blandford} \& {Znajek}(1977)}]{bz77}
{Blandford}, R.~D. \& {Znajek}, R.~L. 1977, \mnras, 179, 433

\bibitem[{Catlett {et~al.}(2007)Catlett, Andrews, Bair,
  {et~al.}}]{catlett2007tao}
Catlett, C., Andrews, P., Bair, R., {et~al.} 2007, HPC and Grids in Action,
  Amsterdam

\bibitem[{{Cenko} {et~al.}(2009)}]{cenko_grb_breaks_2009}
{Cenko}, S.~B. {et~al.} 2009, ArXiv:0905.0690 (astro-ph)

\bibitem[{{De Villiers} {et~al.}(2005){De Villiers}, {Hawley}, {Krolik}, \&
  {Hirose}}]{dhkh05}
{De Villiers}, J.-P., {Hawley}, J.~F., {Krolik}, J.~H., \& {Hirose}, S. 2005,
  \apj, 620, 878

\bibitem[{{Gammie} {et~al.}(2003){Gammie}, {McKinney}, \& {T{\'o}th}}]{gam03}
{Gammie}, C.~F., {McKinney}, J.~C., \& {T{\'o}th}, G. 2003, \apj, 589, 444

\bibitem[{{Koide} {et~al.}(2000){Koide}, {Meier}, {Shibata}, \&
  {Kudoh}}]{koi00}
{Koide}, S., {Meier}, D.~L., {Shibata}, K., \& {Kudoh}, T. 2000, \apj, 536, 668

\bibitem[{{Komissarov}(2001)}]{kom01}
{Komissarov}, S.~S. 2001, \mnras, 326, L41

\bibitem[{{Komissarov} {et~al.}(2007){Komissarov}, {Barkov}, {Vlahakis}, \&
  {K{\"o}nigl}}]{kom07}
{Komissarov}, S.~S., {Barkov}, M.~V., {Vlahakis}, N., \& {K{\"o}nigl}, A. 2007,
  \mnras, 380, 51

\bibitem[{{Komissarov} {et~al.}(2009){Komissarov}, {Vlahakis}, {K{\"o}nigl}, \&
  {Barkov}}]{kom09}
{Komissarov}, S.~S., {Vlahakis}, N., {K{\"o}nigl}, A., \& {Barkov}, M.~V. 2009,
  \mnras, 394, 1182

\bibitem[{{Kumar} \& {Barniol Duran}(2009)}]{kd09}
{Kumar}, P. \& {Barniol Duran}, R. 2009, \mnras, 400, L75

\bibitem[{{Lind} {et~al.}(1989){Lind}, {Payne}, {Meier}, \&
  {Blandford}}]{lind1989}
{Lind}, K.~R., {Payne}, D.~G., {Meier}, D.~L., \& {Blandford}, R.~D. 1989,
  \apj, 344, 89

\bibitem[{{Lyubarsky}(2009{\natexlab{a}})}]{lyub09}
{Lyubarsky}, Y. 2009{\natexlab{a}}, \apj, 698, 1570

\bibitem[{{Lyubarsky}(2009{\natexlab{b}})}]{lyub09b}
{Lyubarsky}, Y.~E. 2009{\natexlab{b}}, \mnras, 1753

\bibitem[{{McKinney}(2005)}]{mck05}
{McKinney}, J.~C. 2005, \apjl, 630, L5

\bibitem[{{McKinney} \& {Gammie}(2004)}]{mck04}
{McKinney}, J.~C. \& {Gammie}, C.~F. 2004, \apj, 611, 977

\bibitem[{{McKinney} \& {Narayan}(2007{\natexlab{a}})}]{mck07a}
{McKinney}, J.~C. \& {Narayan}, R. 2007{\natexlab{a}}, \mnras, 375, 513

\bibitem[{{McKinney} \& {Narayan}(2007{\natexlab{b}})}]{mck07b}
{McKinney}, J.~C. \& {Narayan}, R. 2007{\natexlab{b}}, \mnras, 375, 531

\bibitem[{{Michel}(1969)}]{mic69}
{Michel}, F.~C. 1969, \apj, 158, 727

\bibitem[{{Michel}(1973)}]{mic73}
{Michel}, F.~C. 1973, \apjl, 180, L133

\bibitem[{{Mizuno} {et~al.}(2009){Mizuno}, {Lyubarsky}, {Nishikawa}, \&
  {Hardee}}]{mizuno2009}
{Mizuno}, Y., {Lyubarsky}, Y., {Nishikawa}, K., \& {Hardee}, P.~E. 2009, \apj,
  700, 684

\bibitem[{{Narayan} {et~al.}(2009){Narayan}, {Li}, \& {Tchekhovskoy}}]{nlt09}
{Narayan}, R., {Li}, J., \& {Tchekhovskoy}, A. 2009, \apj, 697, 1681

\bibitem[{{Narayan} \& {McClintock}(2008)}]{nm08}
{Narayan}, R. \& {McClintock}, J.~E. 2008, New Astronomy Review, 51, 733

\bibitem[{{Narayan} {et~al.}(2007){Narayan}, {McKinney}, \& {Farmer}}]{nar07}
{Narayan}, R., {McKinney}, J.~C., \& {Farmer}, A.~J. 2007, \mnras, 375, 548

\bibitem[{{Piran}(2005)}]{pir05}
{Piran}, T. 2005, Reviews of Modern Physics, 76, 1143

\bibitem[{{Sikora} {et~al.}(2007){Sikora}, {Stawarz}, \& {Lasota}}]{ssl07}
{Sikora}, M., {Stawarz}, {\L}., \& {Lasota}, J.-P. 2007, \apj, 658, 815

\bibitem[{{Tchekhovskoy} {et~al.}(2007){Tchekhovskoy}, {McKinney}, \&
  {Narayan}}]{tch07}
{Tchekhovskoy}, A., {McKinney}, J.~C., \& {Narayan}, R. 2007, \mnras, 379, 469

\bibitem[{{Tchekhovskoy} {et~al.}(2008){Tchekhovskoy}, {McKinney}, \&
  {Narayan}}]{tch08}
{Tchekhovskoy}, A., {McKinney}, J.~C., \& {Narayan}, R. 2008, \mnras, 388, 551

\bibitem[{{Tchekhovskoy} {et~al.}(2009{\natexlab{a}}){Tchekhovskoy},
  {McKinney}, \& {Narayan}}]{tch09}
{Tchekhovskoy}, A., {McKinney}, J.~C., \& {Narayan}, R. 2009{\natexlab{a}},
  \apj, 699, 1789

\bibitem[{{Tchekhovskoy} {et~al.}(2009{\natexlab{b}}){Tchekhovskoy}, {Narayan},
  \& {McKinney}}]{tch09c}
{Tchekhovskoy}, A., {Narayan}, R., \& {McKinney}, J.~C. 2009{\natexlab{b}},
  preprint (ArXiv:0911.2228)

\bibitem[{{Tchekhovskoy} {et~al.}(2009{\natexlab{c}}){Tchekhovskoy}, {Narayan},
  \& {McKinney}}]{tch09b}
{Tchekhovskoy}, A., {Narayan}, R., \& {McKinney}, J.~C. 2009{\natexlab{c}},
  preprint (ArXiv:0909.0011)

\bibitem[{{Vlahakis} \& {K{\"o}nigl}(2003{\natexlab{a}})}]{vla03a}
{Vlahakis}, N. \& {K{\"o}nigl}, A. 2003{\natexlab{a}}, \apj, 596, 1080

\bibitem[{{Vlahakis} \& {K{\"o}nigl}(2003{\natexlab{b}})}]{vla03b}
{Vlahakis}, N. \& {K{\"o}nigl}, A. 2003{\natexlab{b}}, \apj, 596, 1104

\end{thebibliography}

\end{document}